\documentclass[manuscript]{emulateapj}
\usepackage{graphicx}
\usepackage{rotating}
\usepackage{mathtools}

\begin{document}
\newcommand{\kms}{\mbox{km~s$^{-1}$}}
\newcommand{\s}{\mbox{$''$}}
\newcommand{\mloss}{\mbox{$\dot{M}$}}
\newcommand{\my}{\mbox{$M_{\odot}$~yr$^{-1}$}}
\newcommand{\ls}{\mbox{$L_{\odot}$}}
\newcommand{\ms}{\mbox{$M_{\odot}$}}
\newcommand{\rs}{\mbox{$R_{\odot}$}}
\newcommand\mdot{$\dot{M}  $}
\newcommand{\lsim}{\raisebox{-.4ex}{$\stackrel{<}{\scriptstyle \sim}$}}

\title{A Pilot Deep Survey for X-Ray Emission from fuvAGB Stars}
\author{R. Sahai\altaffilmark{1},  J. Sanz-Forcada\altaffilmark{2}, C. S{\'a}nchez Contreras\altaffilmark{2}, \& M. Stute\altaffilmark{3}}
\altaffiltext{1}{Jet Propulsion Laboratory, MS\,183-900, California Institute of Technology, Pasadena, CA 91109}
\altaffiltext{2}{Astrobiology Center (CSIC-INTA), ESAC campus, E-28691 Villanueva de la Canada, Madrid, Spain}
\altaffiltext{3}{Institute for Astronomy and Astrophysics, Eberhard Karls Universit\"at T\"ubingen, Auf der Morgenstelle 10, D-72076, T\"ubingen, Germany}

\begin{abstract}
We report the results of a pilot survey for X-ray emission from a newly discovered class of AGB stars with far-ultraviolet excesses  
(fuvAGB stars) using XMM-Newton and Chandra. We detected X-ray emission in 3 of 6 fuvAGB stars observed -- the X-ray fluxes are 
found to vary in a stochastic or quasi-periodic manner on roughly hour-long times-scales, and simultaneous UV observations using the 
Optical Monitor on XMM for these sources show similar variations in the UV flux. 
These data, together with previous studies,  
show that X-ray emission is found only in fuvAGB stars. From modeling the spectra, we find that the observed X-ray luminosities are 
$\sim(0.002-0.2)$\,\ls, and the X-ray emitting plasma temperatures are $\sim(35-160)\times10^6$\,K. The high X-ray temperatures 
argue against the emission arising in stellar coronae, or directly in an accretion shock, 
unless it occurs on a WD companion. However, none of the detected objects is a known WD-symbiotic 
star, suggesting that if WD companions are present, they are relatively cool ($<20,000$\,K). In addition, the 
high X-ray luminosities specifically argue against emission originating in the coronae of main-sequence companions. We discuss several models for the 
X-ray emission and its variability and find that the most likely scenario for the origin of the 
X-ray (and FUV) emission involves accretion activity around a companion star, with confinement by 
strong magnetic fields associated with the companion and/or an accretion disk around it. 
\end{abstract}


\section{Introduction}\label{intro}
Almost all of our current understanding of the late evolutionary stages of ($\sim1-8)$\,\ms~stars is based on single-star models. However, binarity can drastically affect late stellar evolution by (a) cutting
short normal AGB evolution or even preventing stars from reaching the AGB, due to a phase of strong binary
interaction, when the primary was an RGB star, and (b) prolonging post-AGB evolution due to
mass-transfer back onto the primary from a circumstellar disk around the companion (van Winckel 2003).
It has long been argued that binarity is responsible, directly or indirectly, for the dramatic and poorly-understood changes in the
history and geometry of mass loss that occurs in stars as they evolve off the AGB to become PNs. A variety of binary models
(e.g., review by Balick \& Frank 2000) have been proposed, which can lead to the generation of accretion disks and magnetic fields. The latter are likely the underlying physical cause for the highly collimated jets that have been proposed as the primary agents for the formation of bipolar and multipolar PNs (Sahai \& Trauger 1998, Sahai et al. 2011a).

However, observational evidence of binarity in AGB stars is sorely lacking simply because AGB stars are very luminous and variable, invalidating standard techniques for binary detection such as radial-velocity and photometric variations due to a companion star. Sahai et al. (2008: Setal08) 
therefore used an innovative technique of searching for UV emission from AGB stars with GALEX (Morrissey et al. 2007) that exploits the  
favorable secondary-to-primary photospheric flux contrast ratios reached in the UV for companions of spectral type hotter
than about G0 ($T_{eff}$=6000\,K) and luminosity, $L\gtrsim1$\ls.  Setal08 detected emission from  
9/21 objects in the GALEX FUV ($1344-1786$\,\AA, effective $\lambda$=1516\,\AA) and NUV ($1771-2831$\,\AA, effective $\lambda$=2267\,\AA) bands; since these objects (hereafter fuvAGB stars) also showed significant UV variability, 
Setal08 concluded that the UV source was unlikely to be solely a companion's photosphere, and was dominated 
by emission from variable accretion activity.
 
From a subsequent search of the MAST\footnote{Mikulski Archive for Space Telescopes}/GALEX archive, we found about 100 fuvAGB stars $\gtrsim5\sigma$ detections in the FUV band. We required the UV 
source position to be coincident, within $3{''}$, with the optical
position of the AGB target stars (sp.\,types M4 or later). 
The chance coincidence probability with random sources for 
our objects is extremely low (Setal08). Even for the hottest sources in 
our catalog (sp.\,type M4), the detected FUV fluxes ($\gtrsim$20\,$\mu$Jy) correspond to a 
significant excess above photospheric emission -- e.g., for an M4\,III giant 
with $L\sim6000$\,\ls, $T_{eff}\sim3560$\,K (e.g., Ridgway et al. 1980, Perrin et al. 1993), and distance 500 pc, the blackbody 
flux at 1500\,\AA~is $6.2\,\mu$Jy.  Many fuvAGB stars show extreme UV variability as well, 
such as Y\,Gem (Sahai et al. 2011b: Setal11). 

We report here the results of a pilot search for X-ray emission from a small subset of fuvAGB stars. 
Our detection of X-ray emission from 3 objects more than 
doubles the known number of X-ray emitting AGB stars, providing, for the first time, high-quality X-ray spectra and    
X-ray and UV light curves. 

\section{Observation \& Results} \label{sec_obs}

\begin{turnpage}
\begin{figure}[htb]
\vspace{-0.7in}
\hspace{-0.4in}
\resizebox{1.3\textwidth}{!}{\includegraphics{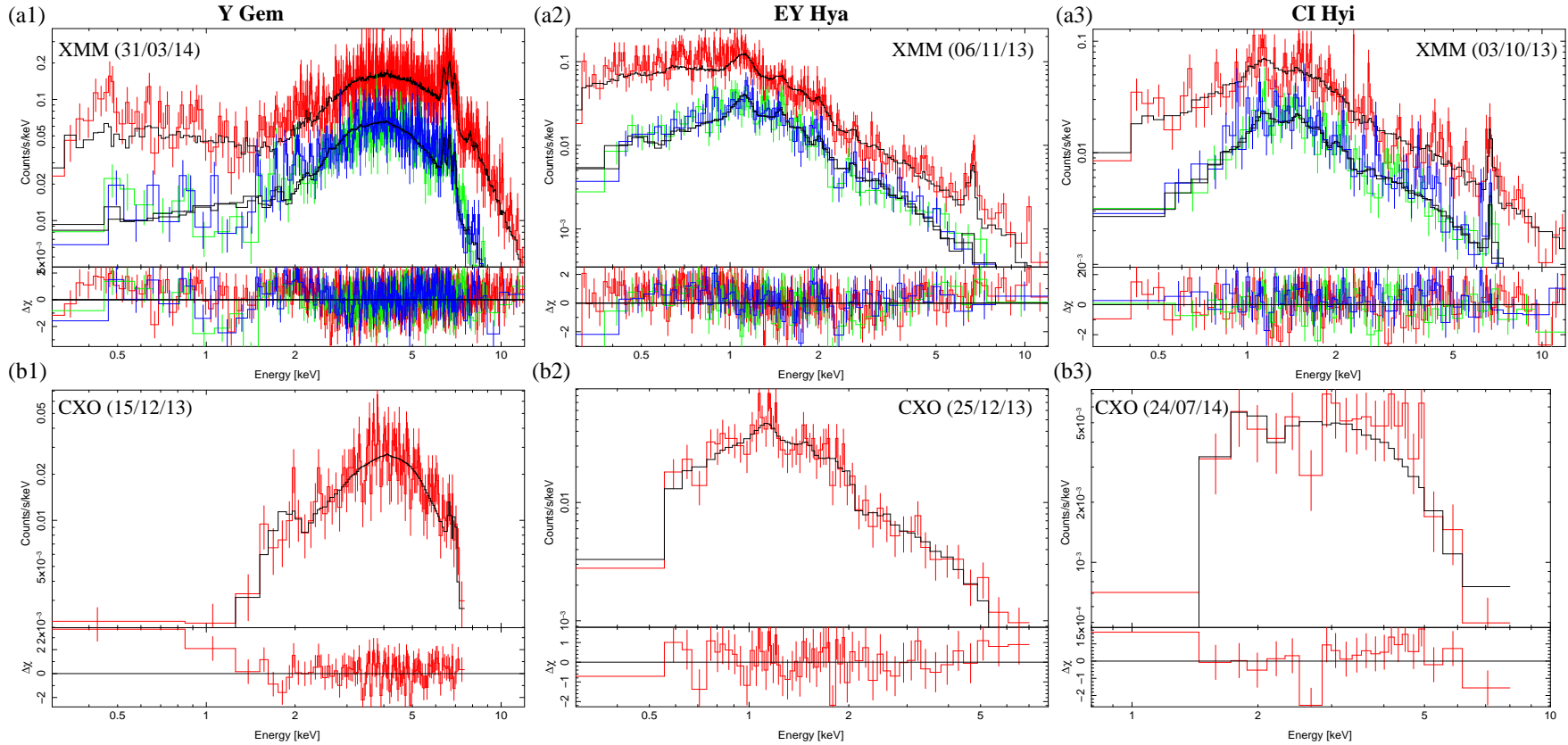}}
\vskip 0.02in
\hspace{-0.4in}
\parbox{8.8in}{\caption{X-ray spectra (colored curves) and model fits (black curves) of the fuvAGB stars Y\,Gem, EY\,Hya and CI\,Hyi. Panels rows show XMM/EPIC 
(pn:\,red, MOS1:\,green, MOS2:\,blue), and CXO (ACIS-S). Spectra shown include the background; the ISIS model fitting procedure takes the latter into account (see text for details).}
\label{xspec}
}
\end{figure}
\end{turnpage}

We observed 5 sources with {\em XMM-Newton} 
using EPIC in full window mode, 
(AO-13  Priority C proposal 072034), and 4 sources  
with the {\em Chandra X-Ray Observatory} (CXO) using ACIS-S (Cycle 15 proposal 15200476.)  
Three sources, CI\,Hyi, Y\,Gem \& EY\,Hya, were common to both programs, and were detected in X-ray emission (Fig.\,\ref{xspec}a,b). Simultaneous UV observations, providing us UV light-curves for these 3 sources were obtained using the Optical Monitor on XMM through the UVM2 ($\lambda=231$\,nm) and UVW2 ($\lambda=$212\,nm) filters in Fast Mode. Essential observing details are given in Table\,\ref{Tbl_obs}. 
Data were reduced following standard procedures, 
and cleaned by removing time intervals affected by high background, and the cleaned observations were used to extract the X-ray spectral energy distributions in the 0.3--10 keV\,range (see, e.g., Sanz-Forcada et al. 2011).

\begin{turnpage}
\begin{figure}[htb]
\vspace{-0.7in}
\hspace{-0.4in}
\resizebox{1.3\textwidth}{!}{\includegraphics{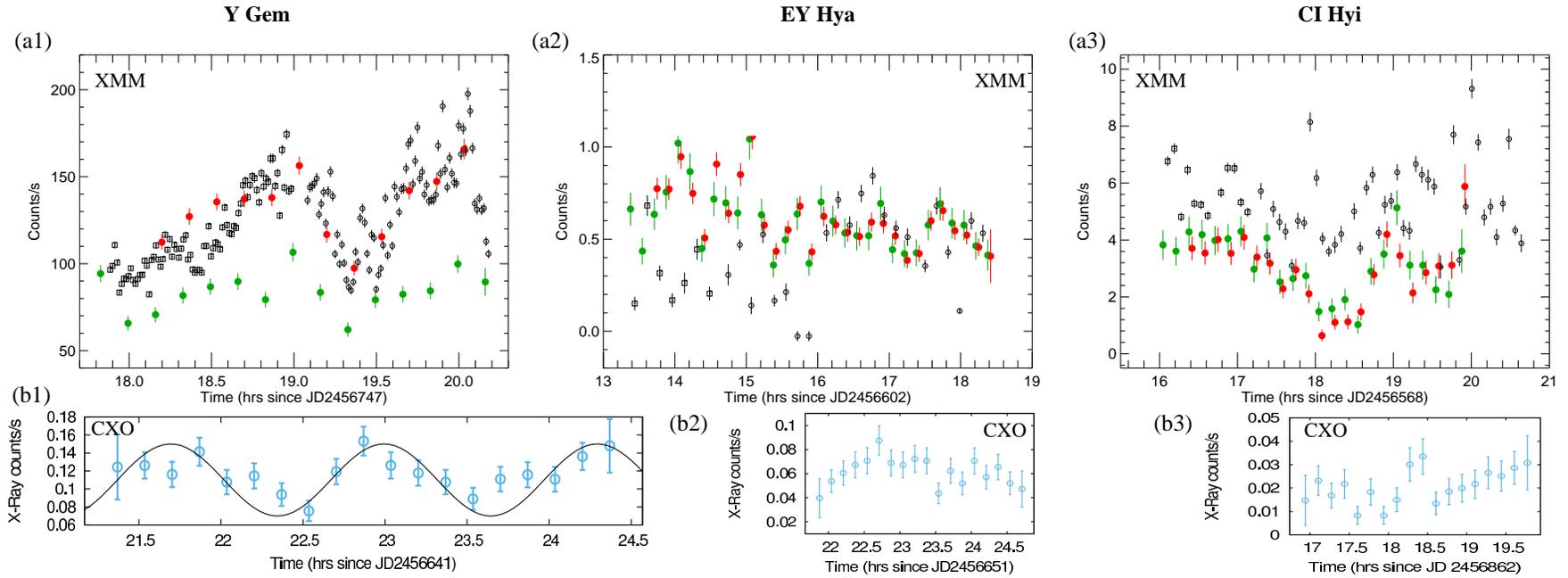}}
\vskip 0.02in
\hspace{-0.4in}
\parbox{8.8in}{\caption{X-ray and UV light-curves of the fuvAGB stars Y\,Gem, EY\,Hya and CI\,Hyi. The energy range used to extract 
the X-ray curves was 0.3-10\,keV for XMM, and 0.3--7\,keV for CXO. Panels show data from XMM (EPIC=pn+MOS1+MOS2:\,red, MOS=MOS1+MOS2:\,green, UVM2:\,black squares, UVW2:\,black circles), and 
CXO (ACIS-S). The EPIC, MOS and UVW2 data have been respectively re-scaled as follows: 80,\,130,\,2.5\,(Y\,Gem), 2,\,5,\,0.8\,(EY\,Hya), 20,\,30,\,2.5\,(CI\,Hyi). A sinusoidal fit (by eye) with period, P=1.35\,hr is shown for Y\,Gem. All data are background-subtracted; error bars are $\pm1\sigma$. In order to facilitate the comparison between the XMM and CXO  light curves, the widths of the two panels for each source have been 
adjusted so that similar time-intervals have similar lengths (on the horzontal axes).}
\label{xuv_lc}
}
\end{figure}
\end{turnpage}

Short-term flux X-ray variations are seen during the course of each observation, spanning a factor $\sim (2-3.5)$ on hour-long timescales (Fig.\,\ref{xuv_lc}a,b). The energy ranges used to extract these curves 
was $0.3-10$\,keV for the XMM/EPIC instruments, and $0.3-7$\,keV in CXO/ACIS (for the latter,
the removal of data for $E>7$\,keV is because that region has more problems due to background noise,
as explained in Chandra documentation.) For CI\,Hyi and EY\,Hya, the variations 
appear to be stochastic or quasi-periodic. In Y\,Gem, the light-curves give stronger evidence of a discrete period of 
about 1.2--1.4\,hr, best seen in the CXO/ACIS-S light curve  (Fig.\,\ref{xuv_lc}b1) that spans a longer period than the EPIC one. For the longer time-scales between the 
XMM and CXO observational epochs for each source, we find changes in the observed fluxes up to a factor 1.8 (in Y\,Gem and EY\,Hya); however a direct comparison cannot be made between the spectra as these were taken with different instruments.

The UV fluxes also show short-term variations like the X-ray fluxes (Fig.\,\ref{xuv_lc}a). Since the UVM2 and UVW2 data were taken sequentially (the OM can take data only through one filter at a time), we have scaled the UVW2 count rate to match that of UVM2 at the time of transition between the two. We find that the UV and X-ray variations appear reasonably well correlated in the case of Y\,Gem and CI\,Hyi. For EY\,Hya, the correlation appears to be weaker.

The ISIS package (Houck \& Denicola 2000) and the Astrophysics Plasma Emission Code (APEC, Smith et al. 2001) were used to fit the spectra with thermal models. The background 
spectrum was estimated from an empty field on the same detector chip as the source, and provided to the ISIS software that then used it to compute and 
fit a model "Source+Background" spectrum to the total spectrum extracted from a circular aperture centered on the source. For XMM, we simultaneously fit the data from the EPIC pn, MOS 1 and MOS 2 detectors. For each object we first fit the EPIC spectrum as it has higher S/N than the ACIS one. 
The elemental abundances used are those of Anders \& Grevesse (1989), except for the Fe abundance ([Fe/H]=log10 of the 
Fe abundance divided by the solar value) which was chosen to be a free parameter only if the S/N was sufficiently high, 
or fixed appropriately (see below).

\begin{figure}[htb]
\rotatebox{270}{\resizebox{0.7\textwidth}{!}{\includegraphics{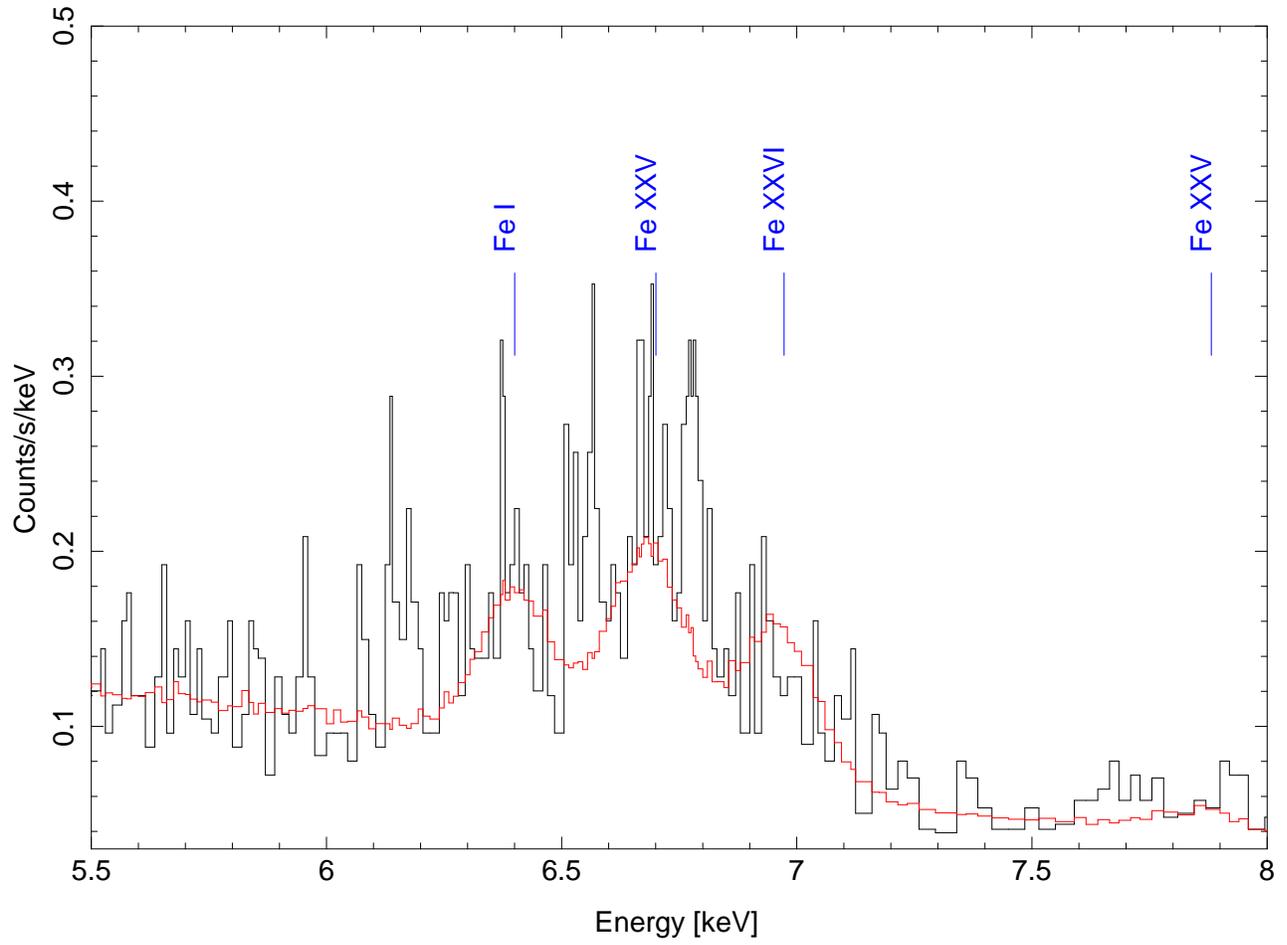}}}
\caption{The Fe line region in the Y\,Gem EPIC/pn spectrum. The expected locations of the Fe\,I line and the line complexes of the coronal Fe\,XXV and Fe\,XXVI lines, 
are marked.
}
\label{ygem-fe}
\end{figure}

All 3 sources also show a weak line feature at $\sim6.3-6.8$\,keV in their XMM spectra, that is likely due to emission from   
(a) the Fe\,XXV,\,XXVI lines complexes at $\sim6.7$\,keV from the very hot plasma that we have found in these objects, and/or 
(b) the 6.4\,keV Fe\,I K$\alpha$ line. We present an expanded version of the Y\,Gem XMM pn spectrum that clearly shows these lines in Figure\,\ref{ygem-fe}. 
The origin of Fe\,I K$\alpha$ line may be the same as in 
Young Stellar Objects (YSOs), where it has been inferred to be flourescent emission from cold, neutral ($<1$\,MK) material in a 
disk irradiated with energetic ($E\gtrsim7.1$\,keV) X-rays (Favata 2005). 

The most robust parameters from our models are the X-ray temperature (Tx) and the emission measure (EM), followed by the hydrogen absorption column density (NH) -- these provide 
us with the intrinsic X-ray luminosity (Lx) of each of our objects. We find that ${\rm Lx}\sim(0.002-0.2)$\,\ls~and ${\rm Tx}\sim(35-160)$\,MK for 
our detected objects (Table\,\ref{Tbl_mod}). 
For the minor Fe\,I line component, our derived values of the line flux are sensitive 
to the emission measure (EM) of the hot plasma that contributes 
to the nearby Fe\,XXV,\,XXVI complex, and are 
less robust than indicated by the formal uncertainties derived from the least-squares fitting. 

The simplest fit to the Y\,Gem XMM spectrum (Epoch 2, Fig.\,\ref{xspec}a1), which has the highest S/N of all the spectra reported here, requires 
two APEC components with the same temperature but with different levels of absorption (NH(1) and NH(2)) and emission measure (EM(1) and EM(2)), 
and a gaussian Fe\,I line. For the latter, we had to 
impose the condition that the line is narrow, and we therefore fixed its width (FWHM) at a nominal value of 0.002\,keV\footnote{Since this width is much less 
than the instrumental resolution, its actual value does not affect the derived line flux}; the line energy was fixed at 6.4\,keV. 
We allow [Fe/H] to vary. 
The Fe\,I line is assumed to be extincted by NH(2), which is significantly smaller than NH(1) -- assuming this line comes from 
a cold disk, the lower attentuation could be an effect of disk inclination and viewing geometry, with the flourescing region on the 
disk having a smaller amount of extincting gas between it and the observer. The derived [Fe/H] 
is not significantly different from the 
solar value. For the CXO spectrum (Epoch 1, Fig.\,\ref{xspec}b1), which has fewer counts, we derived our model fit with two components as above, but without a  
Fe\,I line (as none could be seen in the spectrum), and we fixed 
the iron abundance to the XMM-model value.

For CI\,Hyi and EY\,Hya, we fit the EPIC spectra 
with a 1-T APEC model (Fig.\,\ref{xspec}a2,a3), allowing [Fe/H] to vary. A gaussian Fe\,I line component was also included for fitting EY\,Hya, in the 
same manner as for the modeling of the Y\,Gem XMM spectrum. 
The derived [Fe/H] values are not significantly different from the 
solar value. For the lower S/N ACIS spectrum of EY\,Hya (Fig.\,\ref{xspec}b2), 
we fitted 1-T models fixing the value of [Fe/H] to be the same as in 
the EPIC fit, and derived the required Tx and NH. For the even lower S/N ACIS spectrum of CI\,Hyi (Fig.\,\ref{xspec}b3), 
we fixed both Tx and [Fe/H] to be the same as in the EPIC fit, and derived the required NH.
Our modeling shows significant variations in NH for each source.

These data, together with previous studies (Ramstedt et al. 2012 [RMKV12], Kastner \& Soker 2004 [KS04]), 
show that X-ray emission is found only in fuvAGB stars -- these include 3 objects from our study (CI\,Hyi, EY\,Hya, Y\,Gem) and 2 from archival 
data, R Uma and T Dra (RMKV12). We exclude the fuvAGB star SS\,Vir from this list, since its tentative detection, $4.8\pm1.3$ counts ks$^{-1}$, may be an artifact of optical loading (RMKV12). 
For the set of AGB stars not detected in X-ray emission but with adequate exposure times, $>$0.5\,ks (thus allowing meaningful upper limits) and that were observed with GALEX in the FUV, we have   
2 objects from archival data, RT\,Eri and TX\,Cam (RMKV12, KS04) in addition to the 3 non-detected objects from our study (NU\,Pav, del01\,Aps, and V\,Hya).  
However, as TX\,Cam and RT\,Eri are not fuvAGB stars (we find $3\sigma$ upper limits of 8.5 and 12.4 $\mu$Jy, respectively, for their FUV fluxes), these are not considered 
further.

Amongst the well-known symbiotic systems with AGB primaries, Mira, R\,Aqr and CH\,Cyg, only Mira has been observed in the FUV with GALEX, is both an 
X-ray emitter (Karovska et al. 2005) and a fuvAGB star (Setal11), and believed to have a WD companion (Sokoloski \& Bildsten, 2010). 

\begin{figure}[htb]
\resizebox{0.95\textwidth}{!}{\includegraphics{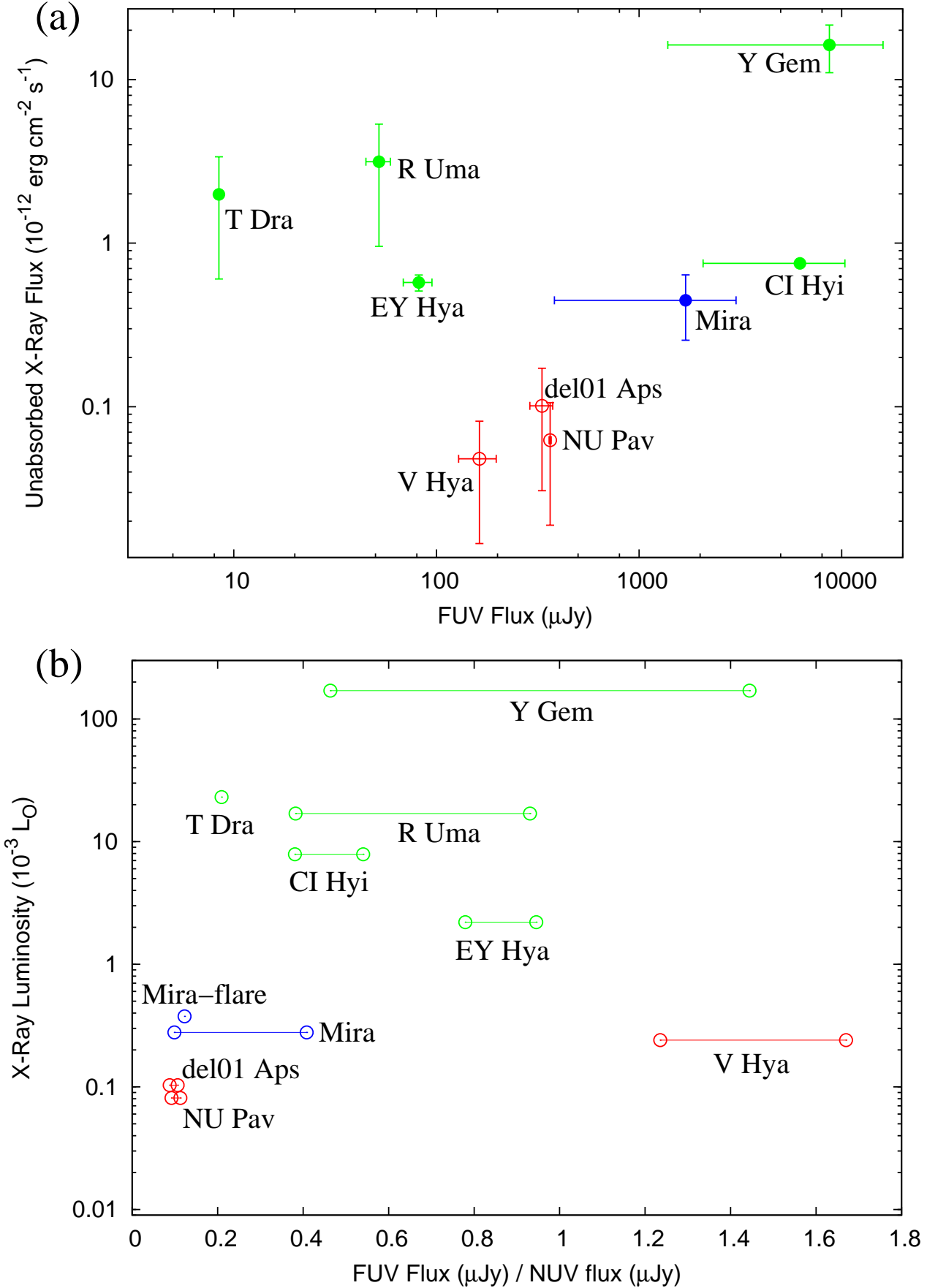}}
\caption{(a) The average (unabsorbed) X-ray flux versus the average (absorbed) FUV flux for fuvAGB stars (green symbols) and Mira, a symbiotic star with an AGB primary (blue symbol). Upper limits ($3\sigma$) are shown for fuvagb stars not detected in X-rays (red open symbols). (b) Average (unabsorbed) X-ray luminosity versus the FUV-to-NUV flux ratio, $R_{fuv/nuv}$, for fuvAGB stars, with green \& blue symbols representing X-ray detections (red symbols show 3$\sigma$ upper limits). In the case of the symbiotic star, Mira, the plot shows its quiescent X-ray luminosity versus the $R_{fuv/nuv}$ range, as well as the X-ray luminosity during a flare on 2003, Dec 6 versus the $R_{fuv/nuv}$ value from a near-contemporaneous epoch. (The error bars typically represent the full range of values when multi-epoch observations are available -- see text for a full explanation of the error bars).
}
\label{fxLxfuv}
\end{figure}

In Fig.\,\ref{fxLxfuv}a, we plot the unabsorbed, average, X-ray flux versus the observed average (GALEX) FUV flux for fuvAGB stars (and Mira) -- it appears that the FUV flux is not a good indicator of the X-ray emission. For the FUV data, the error bars show the flux ranges when data are available for more than one epoch. For the X-ray data, the error bars are as follows : (i) for the 3 detected sources 
in this paper, the errors bars show the range covered by the X-ray fluxes over two epochs, (ii) for the 2 detected sources from RMKV12, the error bars 
show the range in the unabsorbed flux assuming $log(NH)\sim21.5\pm0.4$ (as NH is not well constrained in their models, we assume a typical NH value and a conservatively large uncertainty), (iii) for each of the 3 non-detected sources in this paper, we use the Lx versus NH relationship in Fig.\,4 of RMKV12 for T\,Dra, and $log(NH)\sim21.5\pm0.4$, to 
convert the upper limit on the X-ray count rate to an Lx upper limit. The FUV fluxes have not been corrected for dust extinction (intrinsic or extrinsic to the source) as we do not have a reliable way of estimating these -- however, it is unlikely that 
accounting for these would affect our conclusion above. Since our co-eval 
observations show at least a modest correlation between the variations of the X-ray and UV fluxes, implying a close relationship between the 
X-ray and FUV emission, the observed absence of a correlation in non-coeval observations is likely due to the fact that both are variable.

In Fig.\,\ref{fxLxfuv}b, we plot the average X-ray (i.e., 0.3-10\,kev) luminosity versus the GALEX FUV/NUV flux ratio, $R_{fuv/nuv}$. 
For stars with multi-epoch UV observations, the error bars cover the full range between minimum and maximum values of $R_{fuv/nuv}$.
We find $R_{fuv/nuv}>0.17$ in the 5 fuvAGB stars with X-ray emission (and Mira.) Amongst the 3 fuvAGB stars without 
X-ray emission, $R_{fuv/nuv}<0.12$ for NU\,Pav and del01\,Aps; for V\,Hya, $R_{fuv/nuv}=1.1$. 
Thus $R_{fuv/nuv}$ appears to be a better indicator of X-ray emission than the FUV flux -- but we must caution that this is a tentative 
result since the total number of fuvAGB sources in this plot is small.

\section{Discussion}

We now discuss several models for the X-ray emission that we have observed from fuvAGB stars -- all but one of these require binarity (note that at present there 
is no other evidence of binarity in our targets).

\subsection{X-ray and UV Variability in fuvAGB Stars}
The X-ray and UV variability provide useful constraints on the nature and diagnostics of the emitting region. The observed variability 
is perhaps the strongest indication that these emissions are related to the presence of a binary companion.

Although the short-term variations in EY\,Hya and CI\,Hyi don't show obvious periodicity, the variability time-scales are similar to that 
in Y\,Gem, hence it is plausible that the variability mechanism is the same in all three objects. 

\begin{figure}[htb]
\rotatebox{270}{\resizebox{0.7\textwidth}{!}{\includegraphics{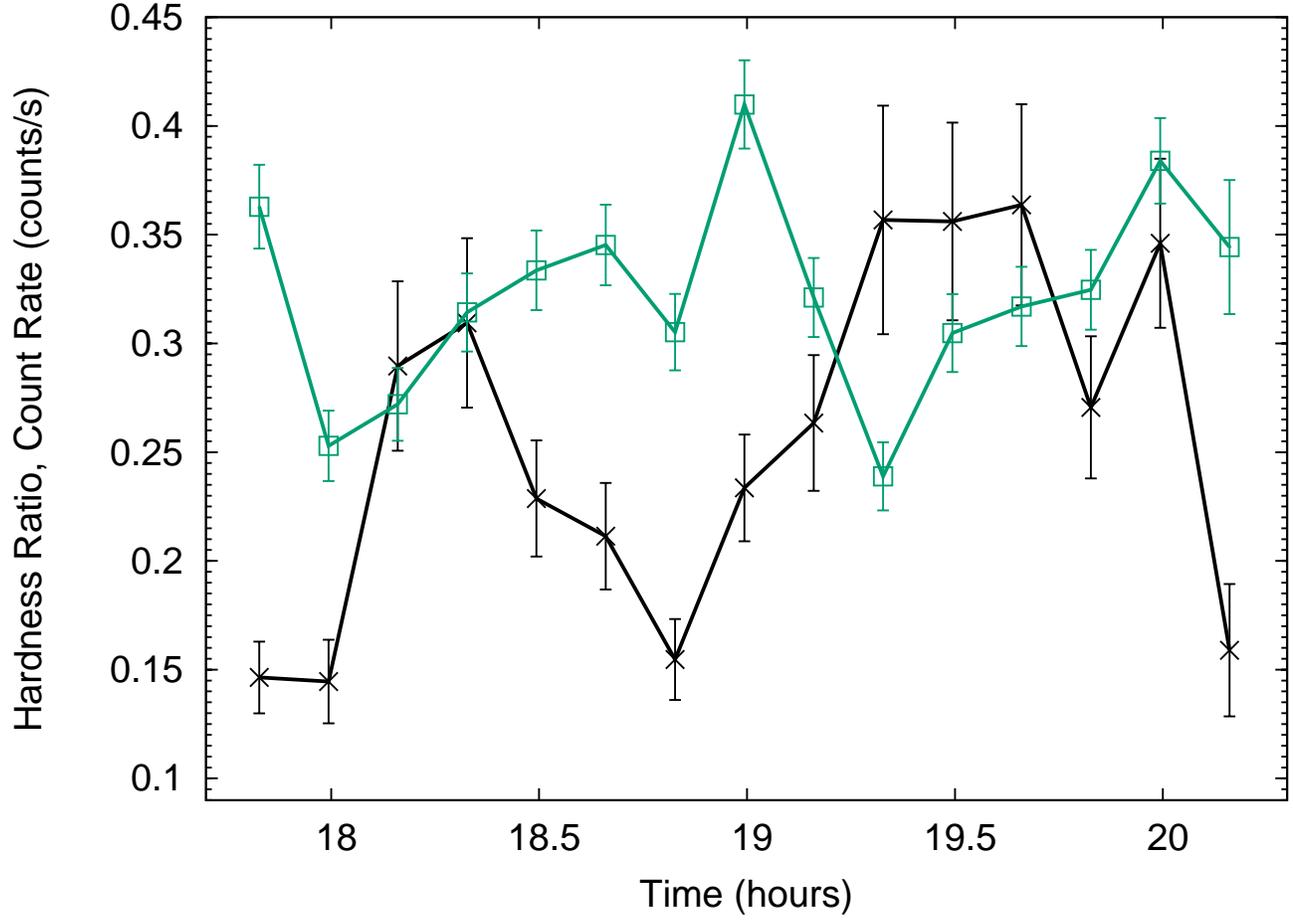}}}
\caption{Light curves of the X-ray EPIC/MOS count-rate (green) and the hardness ratio (black), defined as $(H-S)/(H+S)$, where $H$ and $S$ are the 
fluxes in the energy bands 4--10\,keV and 2--4\,keV, for Y\,Gem (error bars are $\pm1\,\sigma$.)
}
\label{ygem-hard}
\end{figure}

It is unlikely that the X-ray variability is due to flare activity since the fuvAGB stars' X-ray light curves  
do not show the temporal behaviour seen in YSO flares that have rapid rise times
followed by slower exponential decay. Secondly, in flares the hardness ratio increases during the flare peaks, but 
for Y\,Gem, where the XMM count rate is high enough to 
determine the hardness-ratio variations, we find an anti-correlation between these two quantities (Fig.\,\ref{ygem-hard}). We 
defined the hardness ratio as $(H-S)/(H+S)$, where $H$ and $S$ are the fluxes in the energy bands 4--10\,keV and 2--4\,keV. We excluded 
the flux at energies shorter than 2\,keV in defining this ratio because the former has much larger errors due to the smaller number of counts. 

The hardness ratio variations in Y\,Gem are consistent with 
the X-ray changes occuring predominantly as a result of changes in NH (for EY\,Hya and CI\,Hya, the count rates are too low for 
such hardness-ratio light-curve analysis.) We note that our modeling of the X-ray spectra shows that NH varies significantly 
on long time scales in all 3 X-ray fuvAGB stars. It is thus likely that NH variations are the dominant cause of
short and long-term X-ray variations in fuvAGB stars, which may be produced by changes in viewing geometry (e.g.,  due to a warped rotating 
accretion disk), the presence of variable accretion streams, or both.

\subsection{Models for X-ray Emission from fuvAGB Stars}
\subsubsection{Model 1. Accretion Disk around a Companion}
If we assume that the periodicity in Y\,Gem is associated with the orbital period of a central binary, then using Kepler's law, we find 
that the semi-major-axis is, $a\lsim0.0036-0.0046$\,AU or $(5.4-6.8)\times10^{10}$\,cm, assuming primary 
and companion masses of $M_p=1-3$\,\ms, $M_c\lsim1$\,\ms. Since 
such values for $a$ would place the companion deep inside the AGB star's photosphere, the X-ray variation is unlikely to be 
associated with the orbital period of the binary. 

The $\sim$1.3\,hr time-scale is similar to the period of material orbiting close to the inner radius of an 
accretion disk around a sub-solar mass companion, i.e., with $M_c\lsim0.35$\,\ms (implying $a\lsim3\times10^{10}$\,cm); for larger masses, the orbit radius becomes smaller than the stellar radius. 
Y\,Gem's X-ray and UV variability therefore suggests that these emissions arise at or near the magnetospheric radius in a truncated disk, or the boundary layer between the disk and star. In either case, 
the relatively large emission measures derived from our APEC modelling imply large densities, $n$, in the emitting region.  Assuming that the emission comes from a torus of inner radius $r_{in}$ and circular cross-section with fractional radius $\delta\,r$, we find that 
\begin{equation}
 \begin{multlined}
n=2\times10^{11}\,\rm cm^{-3}\,(EM/10^{54}\,\rm cm^{-3})^{1/2}\\
\shoveleft[1cm]{\times(3\times10^{10}\,\rm cm/r_{in})^{3/2}\,(0.2/\delta\,r)^{1/2}}.
 \end{multlined}
\end{equation}
The  magnetic field required to confine such a dense, hot plasma is correspondingly large since the magnetic energy density must be greater than or equal to the thermal energy density. We find that the magnetic field at the inner radius, is,  
\begin{equation}
B_{in}\gtrsim 260\,G\,(n/2\times10^{11}\,\rm cm^{-3})(Tx/10^8\,K). 
\end{equation}

For a truncated disk, the magnetospheric radius, $r_{mag}$, depends on the magnetic 
field of the companion, $B_c$, and the accretion rate (\mdot$_{acc}$) (see Eqn.\,8 of Stute \& Camenzind 2005). 
Assuming the same geometry for the X-ray emitting region as above, and the derived value of $B_{in}$ (260\,G), extrapolated  
to the stellar surface assuming a dipole field, we find that,  
$r_{mag}=2.9\times10^{10}\,\rm cm$, if \mdot$_{acc}\sim10^{-9}$\,\my~and $M_c=0.25$\,\ms. The magnetic field at 
the stellar surface is, $B_{*}=560$\,G. The 
derived value of $r_{mag}$ is relatively insensitive to the unknown accretion rate, since $r_{mag}\propto$ \mdot$_{acc}^{-2/7}$ .

Since the values of $r_{mag}$ and $r_{in}$ are so similar, then given the uncertainties 
in the estimates of both, due to our assumptions on the geometry of the emitting region, the magnetic field geometry, and the accretion rate, 
it is possible that the disk is not truncated, but is connected to the star via a boundary layer.

\subsubsection{Model 2. Accretion Onto a Companion}
The generally accepted model for accretion
from the disk into YSOs (the magnetospheric model) envisions plasma being channeled into
magnetic flux tubes and ramming onto the star at essentially free-fall speed (e.g., Favata 2005). The X-ray temperature of the shocked gas 
resulting from such a process is, $T_{s}=3.44\times10^6\,K (M_c/R_c)$ (from Eqn. 9 of Calvet \& Gullbring 1998), where the companion mass and radius, $M_c$ and $R_c$, are in solar units. Thus, for a Sun-like companion, $T_{acc}$ is significantly lower than observed; for more massive dwarf companions, even upto B5, 
$M_c/R_c\lesssim1.5$. Considering substellar companions, only the most massive brown-dwarfs (BDs) provide a somewhat larger value of $M_c/R_c$, e.g., for a BD with mass 0.1\,\ms, $M_c/R_c=1.5$. 
Only for white-dwarf (WD) companions, with $M_c/R_c\gtrsim10-45$, the observed values 
of Tx ($\sim35-160$\,K) can be easily produced as a result of accretion. 

However, none of our X-ray emitting AGB stars is known to be a symbiotic star with a WD companion. For EY\,Hya, an optical spectrum taken with the Palomar 5m telescope shows no emission lines in the 3889--5436\,\AA~region (K. Findeisen, private communication).
But we cannot rule out the possibility that fuvAGB stars have WD companions that are not hot enough to ionize a detectable amount of gas and 
produce emission lines 
(e.g., the Balmer lines or forbidden lines such as [OI]\,$\lambda\,6300$ or [NII]\,$\lambda\lambda\,6549,6583$). If so, these WD companions 
must have cooled to effective temperatures $T_{eff}<20,000$\,K. However, in the case of Y\,Gem, Setal11 have shown that its 
intense UV emission requires a much larger surface area than that of a WD, hence its UV emission cannot be due to accretion onto a WD.

However, if an adequate mass of accreting gas can be pre-accelerated to high speeds ($v_s$) by magnetic reconnection (e.g., de Gouveia Dal Pino et al. 2014, Hamaguchi et al. 2012), 
then the shock temperature is, using Eqn. 9 of Calvet \& Gullbring (1998), $T_{s}=(3/16\,k_B)\,\mu\,m_H\,v^2_s\sim70\,MK\,(v_s/1500\,\kms)^2$, 
assuming a mean molecular weight, $\mu=1.33$, and is not constrained by the mass-to-radius ratio of the companion.  

Hamaguchi et al. (2012) propose that such ``accretion-induced magnetic reconnection" generates the 
X-ray emission in the YSO V1647\,Ori; the differential rotation between star and its accretion disk shears the 
star's magnetic field, causing the field lines to twist and continuously reconnect.

The quasi-periodic variability in the X-ray light curve of V1647\,Ori, rather similar to what we find for 
the fuvAGB stars, is a by-product of this mechanism due to the production of hot-spots on the rotating central star, where, matter that is accelerated to 
high speeds by the magnetic reconnection collides with the star to produces X-ray emission with high Tx.  It is possible that such a model applies to 
Y\,Gem (and possibly EY\,Hya and CI\,Hyi) as well. 

\subsubsection{Model 3: Stellar Coronal Emission}
X-ray emission from stellar coronae has been extensively studied (e.g., G{\"u}del 2004). The relatively high values of Tx that we have found in our fuvAGB sources argues against the X-ray emission coming from such coronae, which typically show values of Tx in the range $\sim(2-10)$\,MK and rarely as high as the lowest Tx value in our sample, $\sim40$\,MK (e.g., Schmitt et al. 1990). We also note that stellar coronae are usually 
accompanied by stellar chromospheres, and the latter produce strong Ca\,II\,H\,\&\,K emission lines -- however the optical spectrum of EY\,Hya  (K. Findeisen, private communication) 
shows no such emission.
We discuss coronal models in more detail below, and present additional arguments against them as being the likely source of X-ray emission in fuvAGB stars.

\paragraph{(a) From a main-sequence companion}  
``Saturated" coronae around low-mass ($<1$\ms) main sequence stars reach a maximum of Lx/Lc$\sim10^{-3}$ (Pizzolato et al. 2003). 
More massive stars (1.1--1.29\ms) follow a lower ratio, Lx/Lc$\sim10^{-3.9}$. Our observed values of Lx lie in the range $0.002-0.2\ls$. 
Then, if this X-ray luminosity comes from coronal emission from a main sequence companion, 
the implied companion luminosity is Lc$>(2-50)$\,\ls. But since Lc$>2\,\ls$, the companions must have masses $>1.1$\,\ms. 
Thus the implied companion luminosities are even higher, Lc$>$15-500\,\ls, implying masses Mc$\gtrsim$12\ms, which 
is not allowed, since the companion mass must be less than that of the primary AGB star (1--8\,\ms).

\paragraph{(b) From the AGB star} Since plasma at temperatures $\gtrsim 0.5$\,keV cannot be gravitationally confined on the surface of giants and 
supergiants (Rosner, Golub \& Vaiana 1985), the very high values of Tx that we have found imply that the emitting plasma in fuvAGB stars, must be  
magnetically confined (and heated). 
A universal relationship between magnetic flux and the power dissipated through coronal heating is suggested by the linear dependence 
between the total unsigned magnetic flux, $\phi_B$ and Lx for the Sun and active dwarf stars (Pevtsov et al. 2003) over 12 orders of magnitude.
Applying this relationship to our fuvAGB stars, with ${\rm Lx}=(0.92-42)\times10^{31}$\,erg\,s$^{-1}$, we get $\phi_B=(0.92-42)\times10^{27}$\,G\,cm$^{2}$, implying, 
for a AGB star with radius 1\,AU, an average magnetic field $B_{av}=(0.3-15)\,{\rm G}/f_{agb}$, where $f_{agb}$ is the fraction of the AGB's star's surface covered by the  
$B$ field ($f_{agb}\sim1$ for a large-scale field and $f_{agb}<<1$ for localized fields). 
Since we also need to produce substantial amounts of plasma implied by our derived values of the X-ray emission measure, around the fuvAGB stars, 
and since these objects are cool, late-type stars, with three of them known to drive cool, dusty molecular winds 
(EY\,Hya, T\,Dra, R\,Uma), the surface filling factor of the coronal gas (and thus $f_{agb}$) is likely to be relatively small. 

The linear relationship between Lx-$\phi_B$ appears to saturate at the largest values of Lx (i.e., at ${\rm Lx}\sim10^{30}$\,erg\,s$^{-1}$, in 
a region populated by T Tauri stars) 
with 5/6 objects showing Lx values a factor of 10 or more below the linear fit. Hence the value of $B_{av}$ implied by Lx in fuvAGB stars 
may be higher by a factor 10 or more. 
Recent studies show that such values of $B_{av}$ are plausible: e.g., magnetic fields strengths of $0.3-6.9$ ($15.8-1945$)\,G have 
been inferred for the stellar surfaces in 3\,AGB stars 
by extrapolating the field-strength from observations of polarization in the 22\,GHz H$_2$O 
maser line, assuming an $r^{-1}$ ($r^{-3}$) radial variation, appropriate for a toroidal (dipole) field (Leal-Ferreira et al. 2013).

However, a sensitive XMM search for X-ray emission in two AGB stars with known or suspected strong B-fields yielded null detections (KS04), making
coronal emission from the AGB star a less likely candidate for producing the observed X-ray emission.

\section{Concluding Remarks}
In a survey for X-ray emission from 6 AGB stars with far-ultraviolet excesses, we  
detected 3/6 objects with relatively high X-ray luminosities and plasma temperatures; furthermore the X-ray flux 
in each of these is variable on hour-long time-scales. In contrast, a similar sensitive search for X-rays in 2 AGB stars with evidence 
for magnetic fields, was unsuccessful, and these stars show no FUV emission. 
We believe therefore that the most likely scenario for the origin of the 
X-ray (and FUV) emission involves accretion activity around a companion star, with confinement by 
strong magnetic fields associated with the companion and/or an accretion disk around it. However, an extended survey of AGB stars with 
and without UV excesses (including those with evidence of magnetic fields) is needed to help us robustly test this hypothesis.

\acknowledgements
We would like to thank an anonymous referee whose review has led to substantial improvements in an earlier version of this paper. We also thank Neal Turner for helpful discussions. RS's contribution to the research described here was carried out at the Jet Propulsion Laboratory, California Institute of Technology, under a contract with NASA. Financial support was provided by NASA, in part from award GO4-15010Z for Chandra Cycle 15 proposal 15200476 to RS.  CSC's work is partially supported by Spanish MINECO through grants CSD2009-00038, AYA2009-07304, and AYA2012-32032.  JSF acknowledges funding from grant AYA2011-30147-C03-03. Some of the data presented in this paper were obtained from the Mikulski Archive for Space Telescopes (MAST) at STScI. STScI is operated by the Association of Universities for Research in Astronomy, Inc., under NASA contract NAS5-26555. Support for MAST for non-HST data is provided by the NASA Office of Space Science via grant NNX09AF08G and by other grants and contracts.

\clearpage
\onecolumngrid
\begin{table}[!t]
\caption{Observations log}
\label{Tbl_obs}
\begin{tabular}{llllllll}
\hline\hline  
\multicolumn{7}{l}{{\em XMM-Newton} observations} \\
\hline\hline
Target & Camera & Filter & Eff.\,Exp  & V     & Date       & X-Ray    & Spect.\\
       &        &        & Time (sec) & (mag) & dd/mm/yyyy & Emiss.?  & Type\\
\hline
CI Hyi    & pn        & Medium & 7897 & 9.3 & 03/10/2013 & Yes & M6 \\
          & MOS1      & Medium & 13843 &  & \\
          & MOS2      & Medium & 14006 &  &  \\
          & OM        & UVM2   & 4000  &  & \\
          & OM        & UVW2   & 11700  &   & \\
\hline
Y Gem     & pn        & Medium & 6238  & 9.09 & 31/03/2014 & Yes & M8 \\
          & MOS1      & Medium & 8250  &  & \\
          & MOS2      & Medium & 8510  &  & \\
          & OM        & UVM2   & 4000  &  & \\
          & OM        & UVW2   & 4000  &  & \\
\hline
EY Hya    & pn        & Medium & 16839  & 9.37 & 06/11/2013 & Yes & M7 \\
          & MOS1      & Medium & 17284  &  & \\
          & MOS2      & Medium & 17968  &  & \\
          & OM        & UVM2   & 4400   &      & \\
          & OM        & UVW2   & 12800 &  & \\
\hline
del01 Aps & pn        & Thick & 11539   & 4.76 & 18--19/08/2013 & No & M5 \\
          & MOS1      & Thick & 13166   &  & \\
          & MOS2      & Thick & 13172   &  & \\
\hline
NU Pav    & pn        & Thick & 10039   & 5.06 & 26--27/09/2013 & No & M6 \\
          & MOS1      & Thick & 11701   &  & \\
          & MOS2      & Thick & 11672   &  & \\
\hline\hline
\multicolumn{7}{l}{{\em Chandra} observations} \\
\hline\hline
Target & Camera &        & Eff.\,Exp  &  V    & Date       & X-Ray   & Spect. \\
       &        &        & Time (sec) & (mag) & dd/mm/yyyy & Emiss.? & Type \\
\hline
CI Hyi    & ACIS-S  &        & 9838   & 9.3  & 24/07/2014 & Yes & M6 \\
Y Gem     & ACIS-S  &        & 10335  & 9.09 & 15/12/2013 & Yes & M8 \\
EY Hya    & ACIS-S  &        & 9839   & 9.37 & 25/12/2013 & Yes & M7 \\
V Hya     & ACIS-S  &        & 9660   & 9.70 & 18/12/2013 & No  & C \\
\hline
\end{tabular}
\end{table}

\clearpage

\begin{turnpage}
\begin{table}[!t]
\caption{X-Ray Emission Properties of fuvAGB Stars}
\label{Tbl_mod}
\begin{tabular}{llllllllll}
\multicolumn{10}{l}{Detected Sources: Models} \\
\hline
Target        & NH                 & log(Tx)   & log(EM)    & Fx\tablenotemark{1}& Lx\tablenotemark{2} & D\tablenotemark{3} & flux[FeI(6.4)]\tablenotemark{4} & [Fe/H]\tablenotemark{5} & $\chi^2$\tablenotemark{6} \\
    & $10^{22}$\,cm$^{-2}$           & K         & cm$^{-3}$  & erg\,cm$^{-2}$\,s$^{-1}$ & $10^{-3}$\,\ls & kpc      &phot\,s$^{-1}$\,cm$^{-2}$& (dex)   \\
\hline
CI\,Hyi/XMM & 0.33                   & 7.87          & 54.2        & $7.47\times10^{-13}$ & $7.8$     & 0.58       & ...                & 0.12                           & 1.2 \\
range\tablenotemark{7} &$0.05,-0.04$  & $0.04,-0.07$ & $0.03,-0.03$& $\pm8.3\times10^{-16}$  &           &            &                    & $0.13,-0.17$                   & \\
CI\,Hyi/CXO & 2.8          & {\it 7.87}\tablenotemark{8} & 54.2        & $4.06\times10^{-13}$ & $7.9$ &            & ...                & {\it 0.12}\tablenotemark{8}    & 1.2 \\
range\tablenotemark{7} &$0.76,-0.59$  & ...        & $0.08,-0.09$ & $\pm3.5\times10^{-15}$  &            &            &                    &                                & \\
\\
EY\,Hya/XMM & 0.095                   & 7.57          & 53.7        & $5.56\times10^{-13}$ & $2.0$    & 0.35       & $6.7\times10^{-7}$ &  $-0.16$                       & 1.35 \\
range\tablenotemark{7} &$0.009,-0.009$ & $0.05,-0.04$& $0.02,-0.02$ & $\pm3.2\times10^{-16}$ &           &            & $\pm6\times10^{-7}$&  $0.11,-0.13$                  &      \\
EY\,Hya/CXO & 0.05                   & 7.74          & 53.7        & $5.85\times10^{-13}$ & $2.4$     &            & ...                & $-${\it 0.16}\tablenotemark{8} & 0.70 \\
range\tablenotemark{7} &$0.04,-0.04$  &$0.13,-0.11$& $0.04,-0.04$  & $\pm5.8\times10^{-15}$ &            &            &                    &                                &  \\
\\
Y\,Gem/XMM & 8.13\tablenotemark{9}    & 8.1 & 55.6\tablenotemark{9} & $1.05\times10^{-11}$ & $226$     & 0.58       &                    & $-0.10$                         & 1.1 \\
range\tablenotemark{7} &$0.40,-0.38$ &$0.03,-0.04$ &$0.02,-0.02$   & $\pm1.6\times10^{-15}$ &            &            &                    & $0.06,-0.07$                   &  \\
           & 0.043\tablenotemark{10}  & {\it 8.1}\tablenotemark{8} & 53.7\tablenotemark{10}  &             &            &            & $2.4\times10^{-5}$ &                                &  \\
range\tablenotemark{7} &$0.02,-0.015$ &             & $0.05,-0.05$ &                     &            &            &$\pm4.8\times10^{-6}$ &                              &  \\
Y\,Gem/CXO & 15.6\tablenotemark{9}   & 8.2   & 55.2\tablenotemark{9} & $3.1\times10^{-12}$ & $115$       &            & ...                & $-${\it 0.10}\tablenotemark{8}  & 1.2 \\
range\tablenotemark{7} &$16.6,-5.8$ &$0.52,-0.13$ & 0.14,-0.21    & $\pm7.5\times10^{-14}$  &            &            &                    &                                &  \\
           & 3.04\tablenotemark{10} & {\it 8.2}\tablenotemark{8} & 54.6\tablenotemark{10}  &         &            &            &                    &                &  \\
range\tablenotemark{7} &$1.8,-1.5$   &                            & $0.35,-0.49$ &            &         &            &            &                       &   \\
\hline
\multicolumn{10}{l}{Non-Detected Sources}\\
\hline
del01 Aps/XMM &                      &         &              &          & $<0.039$\tablenotemark{11} & 0.23       &            &                           &  \\
NU Pav/XMM    &                      &         &              &          & $<0.030$\tablenotemark{11} & 0.16       &            &                           &  \\
V Hya/CXO     &                      &         &              &          & $<0.090$\tablenotemark{11} & 0.40       &            &                           &  \\
\hline

\tablenotetext{1}{The observed X-ray flux in the 0.03-10 kev range}
\tablenotetext{2}{The intrinsic X-ray luminosity in the 0.03-10 kev range}
\tablenotetext{3}{Distances: from Hipparcos parallax; if unavailable, using MK=-7.6 for late-M semi-regular stars, as in Kahane \& Jura (1994)}
\tablenotetext{4}{Fe\,I 6.4 kev line flux (a Gaussian line-shape was assumed and the line energy and width (FWHM) were fixed at 6.4\,keV and 0.002\,keV during fitting)}
\tablenotetext{5}{logarithm of the ratio of star's Fe abundance to that of the Sun}
\tablenotetext{6}{reduced $\chi^2$ value}
\tablenotetext{7}{+$1\,\sigma$,-$1\,\sigma$ values for NH, log(Tx), log(EM), Fx, and the FeI(6.4) line flux}
\tablenotetext{8}{Parameter value (in italics) was fixed, hence no uncertainties are provided}
\tablenotetext{9}{NH(1), EM(2)}
\tablenotetext{10}{NH(2), EM(2)}
\tablenotetext{11}{$3\sigma$ upper limit}
\end{tabular}
\end{table}
\end{turnpage}

\clearpage

\clearpage

\begin{thebibliography}{}
\bibitem[Anders \& Grevesse(1989)]{ag89} Anders, E., \& Grevesse, N.\ 1989, \gca, 53, 197
\bibitem[Balick \& Frank(2002)]{bf02} Balick, B. \& Frank, A. 2002, ARA\&A, 40, 439
\bibitem[Calvet \& Gullbring(1998)]{cg98} Calvet, N., \& Gullbring, E.\ 1998, \apj, 509, 802
\bibitem[de Gouveia Dal Pino et al.(2014)]{dpino14} de Gouveia Dal Pino, E.~M., Kowal, G.,
\& Lazarian, A.\ 2014, 8th International Conf. of Numerical Modeling of Space Plasma Flows (ASTRONUM 2013), 488, 8
\bibitem[Favata(2004)]{fav04} Favata, F.\ 2005, \memsai, 76, 337
\bibitem[Favata et al.(2005)]{fav05} Favata, F., Flaccomio, E., Reale, F., et al.\ 2005, \apjs, 160, 469
\bibitem[G{\"u}del(2004)]{gud04} G{\"u}del, M.\ 2004, \aapr, 12, 71
\bibitem[Hamaguchi et al.(2012)]{ham12} Hamaguchi, K., Grosso, N., Kastner, J.~H., et al.\ 2012, \apj, 754, 32
\bibitem[Houck \& Denicola(2000)]{hd00} Houck, J.~C., \& Denicola, L.~A.\ 2000, Astronomical Data Analysis Software and Systems IX, 216, 591
\bibitem[Houck (2013)]{h13} Houck, J.~C 2013, ISIS 1.0 Technical Manual, Chandra X-Ray Observatory Center
\bibitem[Kahane \& Jura(1994)]{kj94} Kahane, C., \& Jura, M.\ 1994, \aap, 290, 183
\bibitem[Karovska et al.(2005)]{kar05} Karovska, M., Schlegel, E., Hack, W., Raymond, J.~C., \& Wood, B.~E.\ 2005, \apjl, 623, L137
\bibitem[Kastner \& Soker(2004)]{ks04} Kastner, J.~H., \& Soker, N.\ 2004, \apj, 608, 978  (KS04)
\bibitem[Leal-Ferreira et al.(2013)]{lf13} Leal-Ferreira, M.~L., Vlemmings, W.~H.~T., Kemball, A., \& Amiri, N.\ 2013, \aap, 554, AA134
\bibitem[Morrissey et al.(2007)]{mor07} Morrissey, P., Conrow, T., Barlow, T.~A., et al.\ 2007, \apjs, 173, 682
\bibitem[Naz{\'e} et al.(2011)]{naze11} Naz{\'e}, Y., Broos, P.~S., Oskinova, L., et al.\ 2011, \apjs, 194, 7
\bibitem[Perrin et al.(1998)]{perr98} Perrin, G., Coud{\'e} du Foresto, V., Ridgway, S.~T., et al.\ 1998, \aap, 331, 619
\bibitem[Pevtsov et al.(2003)]{pev03} Pevtsov, A.~A., Fisher, G.~H., Acton, L.~W., et al.\ 2003, \apj, 598, 1387
\bibitem[Pizzolato et al.(2003)]{piz03} Pizzolato, N., Maggio, A., Micela, G., Sciortino, S., \& Ventura, P.\ 2003, \aap, 397, 147
\bibitem[Ramstedt et al.(2012)]{ram12} Ramstedt, S., Montez, R., Kastner, J., \& Vlemmings, W.~H.~T.\ 2012, \aap, 543, A147 (RMKV12)
\bibitem[Ridgway et al.(1980)]{rid80} Ridgway, S.~T., Joyce, R.~R., White, N.~M., \& Wing, R.~F.\ 1980, \apj, 235, 126
\bibitem[Rosner et al.(1985)]{ros85} Rosner, R., Golub, L., \& Vaiana, G.~S.\ 1985, \araa, 23, 413
\bibitem[Sahai \& Trauger(1998)]{st98} Sahai, R. \& Trauger, J.T. 1998, AJ, 116, 1357
\bibitem[Sahai et al.(2008)]{sfgs08} Sahai. R., Findeisen, K., Gil de Paz, A., S{\'a}anchez Contreras, C. 2008, ApJ, 689, 1274 (Setal08)
\bibitem[Sahai et al.(2011)]{smv11} Sahai, R., Morris, M.~R., \& Villar, G.~G.\ 2011a, \aj, 141, 134
\bibitem[Sahai et al.(2011)]{sng11} Sahai, R., Neill, J.~D., Gil de Paz, A., \& S{\'a}nchez Contreras, C.\ 2011b, \apjl, 740, L39 (Setal11)
\bibitem[Sanz-Forcada et al.(2011)]{sf11} Sanz-Forcada, J., Micela, G., Ribas, I., et al.\ 2011, \aap, 532, AA6
\bibitem[Sokoloski and Bildsten 2010]{sb10} Sokoloski, J. \& Bildsten, L. 2010 ApJ, 723, 1188
\bibitem[Schmitt et al.(1990)]{schm90} Schmitt, J.~H.~M.~M., Collura, A., Sciortino, S., et al.\ 1990, \apj, 365, 704 pJ, 723, 1188
\bibitem[Smith et al.(2001)]{sm01} Smith, R.~K., Brickhouse, N.~S., Liedahl, D.~A., \& Raymond, J.~C.\ 2001, \apjl, 556, L91
\bibitem[Stute et al.(2005)]{st05} Stute, M., Camenzind, M., \& Schmid, H.~M.\ 2005, \aap, 429, 209
\bibitem[van Winckel(2003)]{vw03} van Winckel, H.\ 2003, \araa, 41, 391
\end{thebibliography}
\end{document}